# Synchronised Visualisation of Software Process and Product Artefacts: Concept, Design and Prototype Implementation

Mujtaba Alshakhouri, Jim Buchan and Stephen G. MacDonell✉
*SERL, School of Engineering, Computer and Mathematical Sciences*
*Auckland University of Technology*
*Private Bag 92006, Auckland 1142, New Zealand*
malshakh@aut.ac.nz, jim.buchan@aut.ac.nz, stephen.macdonell@aut.ac.nz

**Abstract**

*Context*: Most prior software visualisation (SV) research has focused primarily on making aspects of intangible software product artefacts more evident. While undoubtedly useful, this focus has meant that software process visualisation has received far less attention. *Objective*: This paper presents Conceptual Visualisation, a novel SV approach that builds on the well-known CodeCity metaphor by situating software code artefacts alongside their software development processes, in order to link and synchronise these typically separate components. *Method*: While the majority of prior SV research has focused on re-presenting what is already available in the code (i.e., the implementation) or information derived from it (i.e., various metrics), the presented approach instead makes the design concepts and original developers' intentions – both significant sources of information in terms of software development and maintenance – readily and contextually available in a visualisation environment that tightly integrates the code artefacts with their original functional requirements and development activity. *Results*: Our approach has been implemented in a prototype tool called ScrumCity with the proof of concept being demonstrated using six real-world open source systems. A preliminary case study has further been carried out with real world data. *Conclusion*: Conceptual Visualisation, as implemented in ScrumCity, shows early promise in enabling developers and managers (and potentially other stakeholders) to traverse and explore multiple aspects of software product and process artefacts in a synchronised manner, achieving traceability between the two.

**Keywords:** Software visualisation, Conceptual Visualisation, software process, conceptual design, feature location, traceability, locality.

## 1. INTRODUCTION

In spite of its comparatively young age, the sub-field of Software Visualisation (SV) has advanced rapidly during the past two decades, with a proliferation of new research being published particularly during the last fifteen years. Much of this research has been highly innovative. Its value to the software engineering (SE) community — particularly in promoting comprehension and awareness — has been substantial, and the field has become well established in the research literature [1].

That said, the range of problems explored in recent SV research has been limited. Not unexpectedly, software visualisation research has been concerned with redressing the intangible nature of software, and this has naturally led researchers to direct their attention toward visualising (aspects of) software product artefacts; by comparison, however, far less attention has been afforded to other facets of software development. In fact, it could be argued that the three major categories of SV that have shaped much of existing SV research, namely visualisation of software static structure, visualisation of program runtime behaviour, and visualisation of software evolution [2], have inadvertently played a role in limiting the field's growth and expansion in considering other important aspects of software. It is our particular contention that contemporary visualisation technologies have the potential to also make visible numerous aspects of the *software development process* that are equally disadvantaged by the intangible nature of the end products (i.e., the software artefacts). Extending the benefits of visualisation to important aspects of development is expected to contribute in rendering these technologies as valuable for an even wider range of software practitioners and practices.

Such concerns have been raised by researchers previously. For example, Storey, Čubranić, and German, in their 2005 paper, called for visualisation techniques that provide *activity awareness* [3]. Similarly, Petre and de Quincey highlighted the need to represent *design concepts* in software visualisation approaches in their 2006 paper [4]. Our review of the recent literature (detailed below), however, shows that the software development process is still not being addressed. While there is SV research that highlights some *effects* of the software process on software artefacts, the majority of it is still directed to approaches and techniques for visualising the product. It is our assertion that visualising the software process *in its own right* not only has several potentially beneficial implications for the software industry, but also the wider stakeholder group. For example, from a cognitive perspective, visualising processes *in the context of* software artefact structure, and vice versa, could be important in terms of increasing stakeholder awareness and understanding of *both* the processes and their implemented product artefacts.

In this paper we address this untapped potential of the SV discipline to benefit the software development process by introducing a novel visualisation approach that firstly captures significant aspects of the development process, in our case Scrum, and then tightly integrates and synchronises these with the product artefacts that are created by it. Inspired by the early work of Petre and de Quincey [4], [5], we call this new approach *Conceptual Visualisation*. The approach enables the visual presentation of user requirements as well as developers' design concepts directly 'over' the visualisation scene, seamlessly linked to the actual implementation of those concepts and requirements. It is expected that this new visualisation technique will better inform and support several software tasks and activities, for example, concept location. The approach should also bring software visualisation technologies closer to practitioners through a range of potential applications of use by a variety of software stakeholders. Example applications include: bidirectional requirements traceability, feature (or concept) searching, development progress monitoring, support of group design reasoning, and improved stakeholder communication. Section 6 provides some concrete examples of such usage scenarios. A proof-of-concept prototype tool that implements this new visualisation concept has been developed in order to demonstrate its viability as well as to assess its potential utility for practitioners.

The rest of this paper is structured as follows. Section 2 discusses the status quo of the software development process in the SV literature to situate the research in this paper. Section 3 reflects on prior literature that has emphasized the need to equip visualisation techniques with a more useful, or at least broader, set of information, and in doing so, provides the context for introducing our *Conceptual Visualisation* approach. Section 4 presents our proposed visualisation technique and describes its key technical details along with the major functional features it facilitates. In Section 5 we cover the tool building and a proof of concept demonstration of the new approach. A preliminary evaluation is then presented in Section 6 followed by some reflections on the approach and its implications for practice in Section 7. Some final concluding remarks are made in Section 8.

## 2. SOFTWARE DEVELOPMENT PROCESSES IN SV RESEARCH

As noted in Section 1, the issue of SV research being confined to the consideration of a relatively narrow range of SE aspects has been discussed from different perspectives in a range of prior research. However, it remains evident in recent literature that the response to such discussions seems to have been rather sparse, with the majority of SV research to date being primarily oriented toward re-presenting the product and relatively little addressing the process. To the best of our knowledge, only two forms of SV research address aspects of the software process: a limited research stream on *human activity awareness*, and, more recently, work on the visualisation of *social network analysis*. Interestingly, this dearth of SV research into process appears to be specific to the field of SE: Petre and de Quincey [4] note that in other scientific and engineering disciplines, it is in fact the *development process* that is primarily addressed by visualisation technologies. An explanation for this limited treatment of the software process by the SV community compared to other domains could be that in these other domains it is the process that is intangible, whereas in SE *both the emergent product and its development process* are intangible.

While the application of SV to the software process has been relatively minimal, it has been extensively applied to the software product. Some researchers have gone as far as to suggest that there is currently an abundance of techniques for visualising software artefacts themselves, but that there is a lack of techniques to address other important aspects of software, or in the design of appropriate visual metaphors to incorporate those aspects alongside the visualised artefacts [3]. It remains that both the software static structure and software evolution visualisation categories have rarely considered aspects of the software development process. It is contended here that development processes carry important information that is potentially valuable for various software tasks but that are commonly not documented or have documentation that has no low-effort, straightforward mechanism for software engineers to link to the source code. We address this with the development of the new tool and approach described in this paper. To the best of our knowledge, there is currently no single tool or approach like this that considers the presentation of the software development process as captured by design concepts *in the context of* the software structure, and vice versa. This new approach we describe presents software code artefacts alongside their development processes directly in the visualisation scene.

In one of the first studies to raise the lack of attention to software process in SV, Storey et al. [3] emphasize the need to promote human activity awareness in SV tools, emphasizing its central importance to its practical utility in answering many relevant questions for software stakeholders. After exploring several SV tools, however, they concluded that only a few offered reasonable support for human activity awareness. In the years since, a handful of new approaches have appeared that do indeed attempt to support some forms of activity awareness in their visualisation techniques, most notably: Manhattan [6], StarGate [7], code_swarm [8], and more recently, Replay [9]. Human activity, however, is an *effect* of the actual development process. Soon after the Storey et al. work appeared, Petre and de Quincey [4] signalled that it was the missing *development process* that should be the focal point behind promoting awareness, and referred to the elements of awareness discussed by Storey and colleagues as being only '*subtle*' aspects of software awareness that are a consequence of attention to software change.

Moreover, in examining the literature, it is evident that the tools that have attempted to support activity awareness or have visualised aspects of development have almost all relied primarily on information extracted from IDEs and version control systems. This includes visualisations via heatmaps [10], action graphs [11], and social network analysis [12]. It is similarly evident, however, that the information made available by such tools (and hence the knowledge that could be represented) is limited in nature and is generally only commit-based data. It does not capture design concepts and is confined instead to modification activities. In fact, the authors of the Replay tool have specifically stated that (p.755) *"… the coarse-grained nature of the data stored by commit-based software configuration management systems often makes it challenging for a developer to search for an answer."* Another issue common of these tools is that, apart from the Manhattan tool, they do not present the extracted data in the context of the actual software structure. From a cognitive perspective (discussed in the next section) this additional functionality could play an important role in supporting a range of software tasks [10].

We argue that by finding a mechanism to represent the core aspects of the software development process *within* the context of software structure visualisation, important questions that software engineers, developers, managers, and customers (and possibly other stakeholders) might pose could be more readily answered. This includes almost all the activity awareness questions discussed previously [3], [4] and that fundamentally revolve around **authorship**, **rationale**, **time**, and the **artefacts** themselves. The next section presents and discusses this mechanism.

## 3. CONCEPTUAL VISUALISATION: CONCEPT AND PRODUCT VISUALLY UNIFIED

### 3.1. Capturing and Presenting the Design Concept

In her paper entitled *"Mental Imagery, Visualisation Tools and Team Work"*, Marian Petre reported an empirical investigation involving expert participants from industry and academia that focused on studying the (p.2) *"relationship between expert reasoning about software and software visualisation"* and *"what experts want to use visualisation for"* [5]. She coined the term 'Conceptual Visualisation' to refer to custom-built visualisation tools that the participants had developed previously as supporting tools demanded by their challenging work duties. In relation to our work she reported a particularly important finding: that these experts wanted visualisation tools that contained '*domain knowledge*' and provided organization based on '*conceptual structure*' rather than program structure, but yet were able to (p.10) *"maintain access to the mapping between the two"*. As is evident from the subsequent literature, however, there does not seem to exist any SV tool or approach that attempts to provide such domain knowledge, or that presents the *conceptual* structure of systems.

In 2006, joined by de Quincey, they comprehensively discussed the lack of representation of *design concepts* and *developers' original intentions and rationale* (which together underpin software products) in SV tools, emphasizing their significant role in informing a wide range of software tasks, e.g., adding further functionality, or bug fixing. They suggested that it is more important to visually present the conceptual design in the visualisation rather than merely *re-presenting* the implementation by itself (p.6), *"… because the information most crucial to the programmer – what the program represents, rather than the computer representation of it – is not in the code. At best, the programmer's intentions might be captured in the comment"*.

The problem that Petre and de Quincey were discussing is a fundamental and significant issue in software engineering and it could well be argued that many of the difficult problems facing software engineering are attributable (at least, partly) to it: that is, the physical separation of the end product (i.e., the software code artefacts) from their original design concepts and user requirements (i.e., the specification). It is well known that software can quickly evolve, that developers and team members can change often, and that it is common for documentation to rapidly become out of date. Tracing features of a system back to their original specifications can therefore become very challenging. This can be important when there is a need, for example, to verify if those features have been implemented according to the original (or even amended) user requirements. Moreover, maintenance can be challenging, as developers often have only the source code from which to deduce the intentions of the original developers and the rationale behind the implemented code components before they can contribute further components to the system or maintain the existing code base.

Presenting code components visually is certainly valuable and can support comprehension, as has been demonstrated extensively in the SV literature. It is our contention, however, that *augmenting* that visualisation with an appropriate form of original conceptual design, along with information concerning the processes that created each of those components (and which are normally not available in the code), may be of even more significance and value. In this paper we identify the design concepts, developers' original intentions and rationale behind them—that is, the conceptual design—along with the actual development activities as *core aspects* of the software development process, and refer to them collectively as 'software processes'. Representing these software processes in SV tools in context and alongside the software artefacts is expected to naturally lead to revealing the impact of developers' and teams' activities on the evolving software and make it readily apparent, along with informing several other software tasks. This is likely why Petre and de Quincey [4] referred to the SV tools examined by Storey et al. [3] (which relied on data extracted from version control systems) as providing only 'subtle' aspects of activity awareness, due to the limited nature of the information that can be derived from version control systems (and whose data primarily revolve around *who* changed *what*, and *when*). In contrast, the conceptual design along with the development activities represent the core sources of information that prior SV research on human activity awareness was attempting, but largely failing, to represent by extracting data from version control systems.

### 3.2. The Case for Scrum

Representing the conceptual design in software visualisation is not an immediately straightforward task, mainly because of the difficulty involved in *capturing* the design concepts in a systematic and modular mechanism. Even though Petre and de Quincey [4] importantly highlighted the benefits and advantages of presenting the conceptual design in SV tools, they did not propose any mechanism for it nor suggest how it might be achieved.

In this paper we consider the agile software development methodology Scrum, a popular modern software development framework, as an example to achieve the *capture* and *presentation* of the conceptual design in software visualisation. The Scrum development process is systematic and modular, making it a good candidate to model and then map to the produced software artefacts. The Scrum model relies on the implementation of small elements of functionality represented as an ordered list of Product Backlog Items or PBIs. In Scrum these PBIs are eventually transformed into identifiable software code artefacts, making mappings between the two a viable task. The agile community uses a reasonably standardized and consistent format to describe the different Scrum artefacts (i.e., PBIs, Sprints, and Releases), suitable to build an abstract data model for the automatic collection and presentation of these artefacts[1].

Most importantly, however, the PBIs in essence stand precisely for the actual design concepts in a system. When written well they literally capture and represent the real requirements of users as well as the original intent behind each feature or function implemented in the system. Furthermore, they are the unit of implementation of the various development activities whether they are *new functionalities* being added, *enhancements* to existing features, or *bugs fixes*. In other words, the program should represent the PBIs. They embody the conceptual design that Petre and de Quincey had emphasized as a critical aspect *missing* from software visualisation techniques.

Even though Scrum utilizes a systematic and modular approach to software development, it still suffers the same problem of '*detachment*' between the original high-level design concepts and the final software product artefacts as do other paradigms of software development (albeit, it suffers this problem to a lesser extent). Stated differently, the development processes and the information pertaining to them are still separate from their end products even in Scrum-practicing environments[2], and so are typically not readily available to developers who might be required to inspect, understand and maintain code. Hence, by merging and synchronising these normally detached processes and products, stakeholders could visually examine and reason about individual products (i.e., system code artefacts) contextually and alongside their original design concepts and development processes.

This in fact gives rise to the notion of *Concept Location* (also known as *Feature Location*), which has been reported by some researchers based on experimental results as one of the most practiced software tasks among developers [10]. The term principally refers to the process of finding the part of the source code that implements a specific domain concept (usually in order to complete a particular work task) [10], [11]. Such empirical findings lend further support to the need for mapping design concepts to their implemented software artefacts and presenting them alongside each other.

A last reflection on the relationship between the work presented here and Petre's seminal paper [5] is her report that experts require visualisations that make available *domain knowledge* and that at the same time enable mapping between the *conceptual structure* and actual *program structure*. It is evident at this point that Scrum artefacts, mainly the PBIs, cover the domain knowledge of a system. Also, the visualisation technique that has been developed here (and which is described below) allows precisely for the kind of mapping that those expert participants found to be an important requisite in an SV tool. *Scrum artefacts,* however, not only make available the domain knowledge, but *activities* pertaining to their enactment also carry valuable information that, as is demonstrated below, is potentially able to better inform other software tasks, including some management tasks.

### 3.3. The Role of Software Structure Decomposition
Kuhn et al. [10] reported that the *physical* structural decomposition of software (i.e., the hierarchical structure comprising modules, packages, classes, and so on, or similar components in non-object oriented software) plays an intrinsic and important cognitive role in the way that developers construct their understanding of and build knowledge about the system they are developing or maintaining. The authors conducted a pilot experimental study that involved a visualisation approach called CODEMAP that used a topic-based layout instead of the conventional package-structure layout. The authors were surprised when they found that the participants consistently used the visualisation (p.119) "*as if its layout were based on package structure—even though they were aware of the underlying topic-based layout*" to solve the different comprehension tasks of the experiment. This at first may seem to go against what Petre [5] has reported of the *conceptual* structure being described as more important by the expert participants in her investigation. However, the fact that a provision of access to the mapping between the two has also been marked as essential in Petre's study, might give a clue to the contention that logical and physical software structures are likely to *both* play a significant and important role for stakeholders in building a mental map of software and to complete various software tasks.

Earlier studies on software cognition theories and comprehension models, such as the highly cited papers of Storey et al. [12] and von Mayrhauser and Vans [13], seem to also support this inference. Many of the comprehension models put forward and discussed in those and other similar papers, such as the top-down model, rely principally on the notion that developers intuitively (and often sub-consciously) build an internal map of the software structure decomposition that the brain then utilizes to recall and locate an artefact of interest. Indeed, some of these studies [13] have equally considered conceptual (or functional) as well as physical structure in their mental models of software cognition.

---

[1] In this paper, we use the terms Features and PBIs interchangeably. We also use the term 'Scrum Artefacts' to refer mainly to *Features, Sprints, Releases* and the *data pertaining to their enactment* as collected by typical agile development management tools.

[2] This is true even for teams who use some popular agile management tools. As part of this work, most of those popular tools were surveyed and none provided any mechanism to link the PBIs to their code artefact manifestations. The best that some tools provided (e.g. Taiga) was to offer a connection to *commit actions* in version control systems.

Moreover, the importance of software structure decomposition, whether conceptual or physical, is in strong alignment with studies on '*spatial memory*', which refers to how the brain utilizes locality to retrieve and recall information [1], [10], [14]. Spatial memory is in fact an important concept that underpins software visualisation theory; by providing spatial representations of software artefacts, users can more readily leverage their natural cognitive abilities.

The work here builds specifically on the ideas just discussed and intentionally takes advantage of these phenomena to propose a visualisation technique that merges and synchronises the conceptual design (represented by Scrum artefacts and their enactment activities) with software product artefacts, and provides visual bidirectional mapping between the two in a single visualisation scene.

### 3.4. Traceability Research

The sections above have considered research that motivates the benefit of close integration between software design concepts in general and the final product artefacts (represented by software code artefacts in particular). A particular sub-field that shares common ground with this work is software traceability research, specifically that which aims to connect some aspects of software design, architecture, development activity, or requirements to software code artefacts. In their 2014 paper [15], Lungu et al. have utilised software structure visualisation to support the activity of software architecture recovery using a tool called *Softwarenaut*. Their work is similar to ours in the fact that both utilise visualisations of the hierarchical decomposition of a system to achieve a form of traceability from product to original design. However, their work is focused on systems architecture recovery with the goal of preventing software architectural erosion, while our goal is primarily focused on connecting software requirements (represented particularly by system functional features) back to software code artefacts. In other words, they capture the architecture while we capture the functional requirements. Both works also share an aspect of technology, which is the open-source FAMIX metal-model used to create and store an abstract representation of the hierarchical decompositions of systems, which in turn facilitates the creation of the corresponding visual glyphs.

Similarly, Mirakhorli and Cleland-Huang [16] seek to bridge the gap between software code artefacts and their original architecture patterns (*referred to as tactics by the authors*) with the intent of preventing architecture degradation during maintenance activities. The authors created an Eclipse-based tool called *Archie* that can traverse code and automatically detect popular architectural patterns and then helps to create trace links between visual models of those patterns and the parts of code that fulfil/relate to each pattern. Once the links are created, the tool enables architects to visually monitor those patterns to prevent their degradation whenever the related code is modified by developers (who also get automatic notifications as a deterring mechanism). Significant effort was expended to *create* and *train* automatic classifiers *(machine learning algorithms)* for the top ten architectural patterns as identified by the authors. The automatic detection and identification is thus limited to those predefined ten architecture patterns (while other patterns can be supported with manual mapping, unless an architect is willing to create and train a corresponding new classifier). This work differs from ours in two primary aspects; first, its problem domain is mainly architecture reconstruction whereas ours is functional requirement traceability; second, ours is based on visualisation of the software physical structure to benefit from its cognitive advantage whereas this work visualises models of architectural patterns. Both works, however, share the intent of connecting code artefacts to some form of software design concepts, and both are based on the Eclipse platform.

Both works above focused on connecting code artefacts to some form of architecture modelling. In contrast, Bohnet and Döllner [17] presented an approach called Software Maps through which they projected code-related quality metrics together with certain forms of development activities on a CodeCity-like visualisation in an attempt to support managers in their decision making processes. For instance, the approach makes visible potential hotspots where refactoring is likely to be beneficial. While representing quality metrics is a widely covered aspect in SV research, the concept of putting these metrics in context with live and time-ranged development activities, such as how many developers are working on a given complex class or what class is being most frequently changed, brings new insights to the field. This research is similar to ours in that it builds on top of Wettel's CodeCity metaphor in seeking to create traceability between code artefacts and some form of development activities. However, the two works differ in the nature of the development activities being treated; they cover modification activities in terms of *who*, *what*, and *when*, while we cover development activities in terms of what code artefact implements what feature. Most importantly, the primary intents are different – their goal is to bring visibility and context to internal software quality for more effective communication and decision making, whereas our goal is to bring software conceptual design in context with code artefacts, hence providing bidirectional traceability between functional requirements and their implementation.

Another form of traceability research is addressed in the well-known sub-field that aims to specifically create trace links between code artefacts and their original software requirements. This particular field is strongly relevant to ours despite the fact that it normally does not address the problem in the context of software visualisation[3] (at least to the extent that we could determine in our literature review). We thus now discuss some of the research works that have dealt specifically with *requirement traceability* regardless of the context of the treatment.

---
[3] It is worth highlighting that we are specifically referring to '*visualisations of requirement traceability*' and not '*runtime trace visualisations*', as there are a good number of works on the later area.

In a study concerned with artefact traceability, Fasano and Oliveto [18] discussed this challenging issue in software management, highlighting that both project planning tools and Process Support Systems (PSSs) are often missing adequate support for enabling artefact traceability, making management of changes a difficult task. In their paper, they introduced a novel software management tool named ADAMS to support fine-grained traceability between software artefacts and the software processes that produced them. In their own words (p.149), ADAMS "*enables the definition of a process in terms of the artefacts to be produced and the relations among them, supporting a more agile software process management than activity-based PSSs*". The concept behind their tool was based on a product-oriented work breakdown structure (WBS) augmented with extra process information to enable a manager to define a hierarchy of software artefacts that each team member would be responsible for creating. It also allowed for relations and dependencies to be defined between the various entities. Each defined WBS entity could then be linked directly to its actual code implementation. The principal objective of ADAMS is very similar to our own and both in fact share the same fine granularity level of mapping between software artefacts and their processes; however, we contend that the solution proposed here has some significant advantages.

A key distinction between the two is that the solution introduced here is based on spatial visualisation that allows for the relationships between the *processes* and the artefacts to be explicitly visible in context with the software structure. The second important distinction relates to the process model being utilized. The WBS framework is becoming less commonly used in the software industry in light of the increasing popularity and growing dominance of agile approaches. Even if it were used, however, the product-oriented WBS is typically defined as a *high-level architectural decomposition* of a project—that is, in terms of functional modules and components—never reaching the fine-grained granularity anticipated by ADAMS. The approach introduced here, on the other hand, is based on the activities and artefacts of the *feature-centric* Scrum methodology, which is a highly popular practice in the community.

While ADAMS did not utilise visualisation, another more recent study that is similarly aimed at supporting project managers through the provision of visual representations is that authored by Jaber, Sharif, and Liu [19]. They recognised the cognitive and interactivity advantage of 3D visualisations and so produced a tool called *3DProjView* that presents the relationships between a project's tasks and its resources in an innovative 3D method. Their study showed significant advantages for 3DProjView over traditional project management representations in terms of management efficiency and effectiveness. However, 3DProjView is mainly aimed at project managers as opposed to also being relevant for developers and, while it presents relationships and information over time of a project's tasks and its human resources, it does not provide any means to create trace links to the software product artefacts – a key element of our work.

While the above works share a range of similarities to varying degrees with the approach we present here, it is clear that the most relevant research that our literature review has revealed is the relatively recent work of Delater, Paech, and Narayan. In their 2012 paper [20], which was augmented by an empirical study a year later [21], they presented a *Traceability Information Model* for the purpose of enabling the collection of requirements-to-source code traceability links. Interestingly, they reported that collection of these links *during* the development process as opposed to at later stages renders it more effective and better enables developers to make use of this information while development progresses. Further, their model places the requirements and integrates them with their relevant artefacts from the project management process, principally acknowledging the benefits of making such aspects of the development process synchronised and contextually available with code artefacts. While not identifying the Scrum development process *per se* behind their information model, their approach does use the concepts of *Feature*, *Work Item*, and *Sprint* to organise the collected requirements. In this regard, their work shares a number of similarities with ours with respect to the principal concept, but differs in the fact that it is not based on visualisation, which we have argued above as bringing particular cognitive advantage.

## 4. VISUALISATION TECHNIQUE

The principal intent of the Conceptual Visualisation approach, as derived from the previous discussions, is to project the software development process in the form of Scrum artefacts and activities over a visual representation of software code artefacts in a synchronised mechanism. While a number of candidate visualisation methods could be used, a 3D visual metaphor has been selected in this work. This is based on the well-documented advantages of this metaphor in other SV application [1], [19], [20]. The other key requirement is to have a well-designed data model for Scrum artefact collection and representation. Next, we provide comment on relevant issues in 3D software visualisation. We then introduce the specific metaphor employed in this work and describe our criteria underpinning its selection. The Scrum artefact capture and representation mechanism is then described, and this is followed by a description of the mapping technique between the visual metaphor and the Scrum artefacts.

### 4.1. Visual Metaphor Selection
The primary characteristics considered in the selection of a suitable visual metaphor are as follows: provision of a clear and non-cluttered mapping of the static structure of software; evidence of its cognitive advantage in aiding comprehension; naturally expressive, and that with the Scrum data projected on it would not appear overloaded; and finally, the promotion of simplicity. After a thorough survey of existing 3D metaphors of software static structure, the **City Metaphor** version of Wettel and Lanza (popularly known as CodeCity) [22] was selected (with some slight enhancements being introduced to it for the purposes of our work). It met all the requirements we were seeking, including empirical evidence of its effectiveness in supporting comprehension [23], [24]. Moreover, the metaphor has proved to be highly versatile (i.e., it can be used to highlight different aspects of software) and as such has seen growing popularity among the SV research community. This treemap-based metaphor and variations of it are also occasionally referred to as Software Maps [29].

Other metaphors that were considered included the Software Landscape metaphor [25] and the Evo-Streets approach [26]. The former was considered to be more complex but to have less expressive power (particularly in terms of providing a global overview of system structure), and did not have empirical evidence of its support for user comprehension. The latter approach has promise but its layout was found to result in much larger city landscapes than the Wettel and Lanza approach which might hinder navigability in a 3D environment.

**4.2. The Suitability of Scrum Providing Modular and Discrete Units of Requirements**

In Section 3.2 we argued in support of the candidacy of the Scrum development methodology for providing a traceable mapping between user requirements and implementation code artefacts. In order to describe how the Scrum model is mapped to the code artefacts (and acknowledging that Scrum is already widely understood), it may be helpful to outline very briefly how the methodology is employed in practice. In the Scrum methodology, system development is carried out through the rapid implementation and delivery of self-contained units of user requirements known generally as *Features or Product Backlog Items (PBIs)*. This occurs in short time-boxed cycles known as *Sprints*, where each sprint contains a pre-determined set of PBIs that collectively comprise an *Increment* (that is potentially releasable). Each PBI normally describes a small functional component of a system (hence the other name, *Feature*) that requires a day or less of working effort to develop. A sequence of sprints (normally fewer than 15 and potentially as few as a single sprint) constitutes a *Release*, which is intended to provide a coherent set of working functionality—a deliverable. A complete system is realised and sustained over multiple iterations of releases.

Given even this brief description, it should be evident that in this context *features or (PBIs)* are the smallest units of user requirements, whose implementation results in the creation of the different system artefacts; which on a similar scale map to classes and methods in an object-oriented software context. Packages are simply logical groupings of classes, but they do not represent immediate manifestations of *features*. Since a feature by definition captures a small and specific functionality of a system, it is thus expected to contribute to the system with a constrained set of new classes or methods, or with additions/amendments to existing classes. In other words, the different system artefacts created are nothing more and nothing less than manifestations of *features*. So, in principle, the main task becomes mapping the Scrum *features* to their related classes and methods. This is the main concept underpinning the introduced visualisation technique.

**4.3. A Scrum Data Model**

For the purpose of the proposed visualisation technique a Scrum data object model is required to facilitate the collection and storage of the Scrum artefacts data. However, there does not seem to exist any official or published standardized format for representing the Scrum data model despite its high profile in the software community and its systematic and modular scheme as used by the practicing agile community—presumably, because the data is typically not intended to be exchanged. Therefore, an XML schema for representing Scrum artefacts and relevant data of their enactment activities was designed and developed for the purpose of this work. The schema was designed to reflect the general scheme that was found to be most commonly practiced by agile communities. To inform this development, data presentation models behind some popular Scrum management tools, including *Taiga*, *Trello*, *OnTime* and *ScrumDesk* ®, were carefully inspected.

A matching logical data object model was also developed as part of the prototype tool, and is used to represent the Scrum data internally after parsing the XML files. Fig. [1a] shows a simplified UML diagram of the Scrum data model (only key data elements are shown – the complete XML schema file is available, however, and can be obtained by contacting the authors).

The central entity of the Scrum Data Object Model is the *Feature* object. It may have an aggregate of *Work Entry* data objects, and itself aggregates to form an individual *Sprint* object, which in turn can aggregate to comprise a single *Release* object. The Feature object constitutes the lowest granularity in the Scrum Data Object Model which serves to capture discrete units of design concepts (in the form of functional requirements) and their related development activities (represented by Work Entries). Each Feature object has the following key attributes; **MethodRefs**: a list of unique identifiers (QNames) pointing to individual methods that are either created or modified in fulfilment of this Feature, **ClassRefs**: a list of unique identifiers (QNames) pointing to individual classes that are either created or modified in fulfilment of this Feature, **WorkEntries**: a list of Work Entry data objects that each capture an *identifier* (*QName* pointing to a method or a class), a *Date*, work *Hours*, and a *Type*, **Tasks**: a collection of tasks that comprise this *Feature* (if applicable), **Category**: designating the feature as either a *New Feature*, a *Bug Fix*, or an *Enhancement*, and finally a **Priority** attribute (which is self-explanatory).

**4.4. Mapping Technique**

*4.4.1. Scrum-to-code artefacts mapping*

**Feature Mapping**. To achieve the mapping between individual features and their manifestation as code artefacts (mainly classes and methods), the unique identifier referred to as the Qualified Name (***QName***) of each code artefact is used as a URI reference. Since each developer is assigned to or selects a set of features to implement, it is thus expected that they would know the classes or methods they have created or significantly modified in their realisation of a particular feature. It is hence presumed in this work that each developer plays a key role in constructing this mapping: the proposed mapping requires that Scrum developers identify the *classes* and *methods* that have been either created or modified as a direct result of implementing a given Scrum feature. At first this might seem impractical, but given the existence and increasing adoption of development management tools by the agile community (e.g., the *Axosoft* official site reports "over 11,000" teams to be using their product), and given the significant advantages and implications as are demonstrated below, such a presumption is considered to be both rational and practical. Developers who use such tools are typically already required to regularly update the status and record the work progress of their features' cards on the dashboards provided by the aforementioned tools. It is thus not far-reaching to additionally require these personnel to designate the affected code artefacts using their QName identifiers (argued further in Section 7.2) when updating their *Work Entries* (see *Work Progress Mapping*). Furthermore, given the small size and short-time requirement nature of Scrum features, each feature is expected to manifest in a limited number of system artefacts.

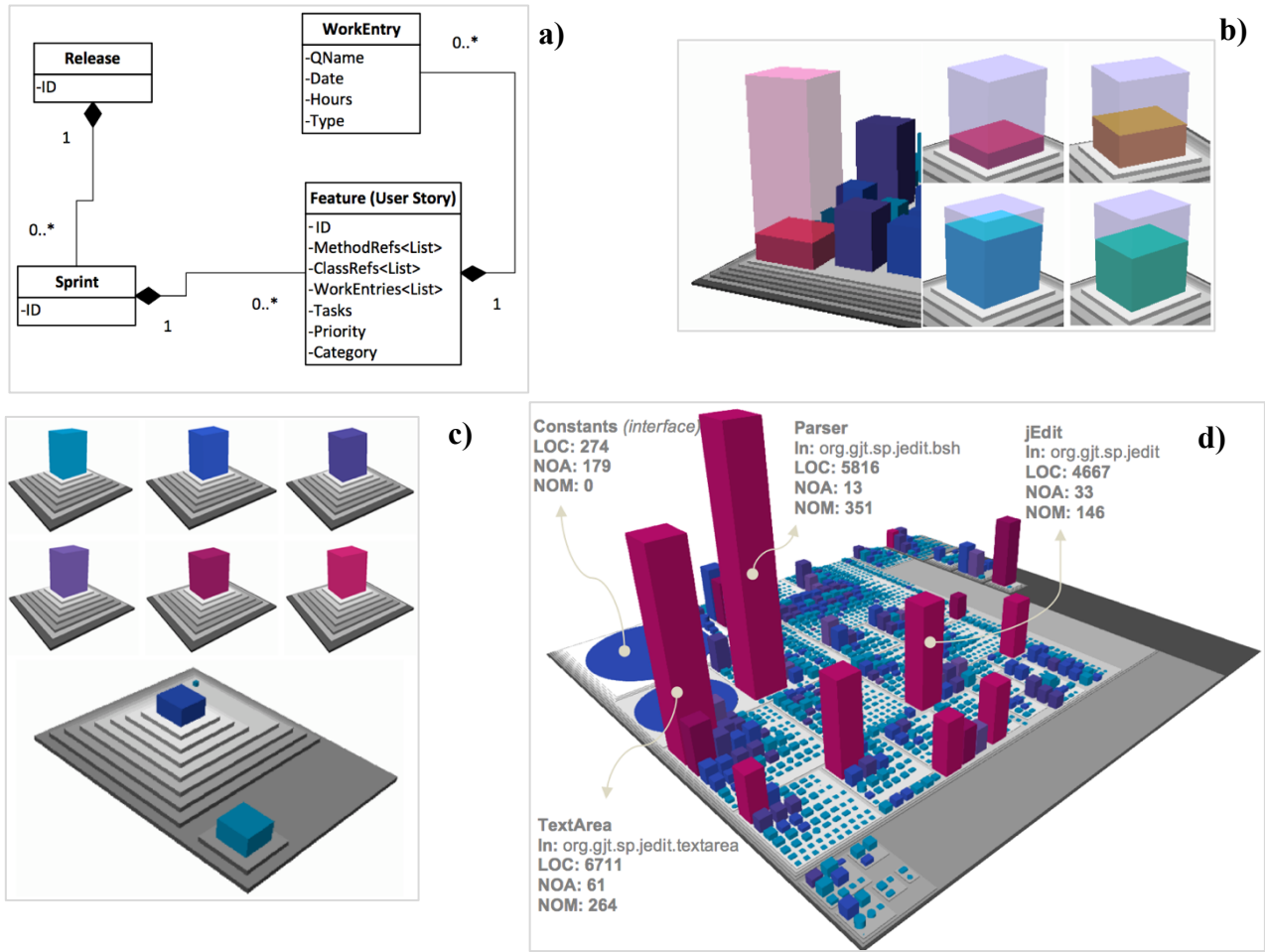

Fig. 1. Clockwise from top left: **a)** A simplified Scrum Data Object Model (only key attributes are shown), **b)** Percentage-to-colour mapping for RC View showing the four configurations of <20%, <40%, <70%, and >70% (only applicable to RC View), **c)** LOC colour mapping showing six configurations of <200, <500, <1000, <1500, and >2000 (applicable in standard view) with colour mapping of package nesting shown below, **d)** City Landscape of jEdit with some mappings identified.

Once features have been designated with the identifiers of their code artefact manifestations, traceability will have essentially been established between the design concepts (i.e., the requirements) and the code artefacts. Aided by the Scrum Data Object Model discussed above, this traceability can be further aggregated and offered to the user at the Sprint level, or even at the Release level (see Fig. [1a]). Furthermore, traceability now becomes achievable and realisable in both directions – from Scrum artefacts to code artefacts, as well as from code artefacts to their corresponding Scrum artefacts (see section 5.3.1).

**Work Progress Mapping**. Another data element key to the proposed visualisation mechanism is *Work Effort*. A concept central to Scrum practice is that of the forecasting of the probable work effort, where each feature is assigned a figure indicating its relative size (perhaps in story points) but that also maps to the *effort* likely to be required for the implementation of that user story. Work Progress is conventionally updated daily by developers (*usually via dashboards provided in management tools*) to reflect how much work is 'remaining', until this figure reaches zero indicating the completion of that feature. This information is captured in our data model by the '**Work Entries**' objects. This data plays an important role in Scrum management tools where it would inform project data reports and predictions, such as burn-down charts. In the proposed visualisation technique, this data element further forms a key data source in realizing a novel multi-granularity visualisation of development progress called '*Remaining/Completed Work View*' (or *RC View* for short) where users can see an immediate *visual* impression of how much work has been completed and how much is left, at three levels; *Feature-level*, *Sprint-level*, and *Release-level*. Furthermore, at *Release* level two mapping scales are offered: users have the option to display the RC view for the entire city (i.e., the entire system, see Fig [3e]) or for individual buildings (classes), the latter offering a more focused view. While the RC View offers three mapping levels at the design concept side (relating to functional requirements), for practical reasons those mappings are always projected onto the class level only at the code artefact side (i.e., we do not go to the method level).

This mapping is enabled using *work entry records* (see the Scrum Data Object Model) which are collected and maintained for each Feature object as developers continue working on them. Work entries are linked to code artefacts using the QName identifier, and are designated as of two types; *Remaining* and *Completed*. In this manner, a trace link can be realised *at work entry level* between the code artefact and its Scrum features. The traceability capability in ScrumCity can thus also be achieved by the QName identifiers that are available at this lower level of the model (in case the *ClassRefs* and *MethodRefs* lists were not populated for instance). As mentioned above, this work presumes a semi-automatic collection of those work entry records where each developer's input is required (see section 4.5).

With the realisation of this mapping, users can monitor the development progress from two perspectives. **Perspective 1**. A user can choose an individual class and view it in RC View mode, or they can expand their selection to an entire package – in which case a visual impression emerges depicting work progress for a particular system component or module. In this perspective, which is initiated from code artefacts side, development progress is calculated with respect to all available work entries for each individual class in the selection. The remaining work-hours of all features related to a class are summed to get the total remaining hours standing for this class. The same is done for the completed work-hours. The ratio of these two values is then computed and the Class glyph is turned partially transparent, where the height of the transparent portion is determined by the percentage of the remaining work. Furthermore, the portion representing the completed work is colour-coded (see Fig. [1b]) to provide a richer visual sensation of development progress. For example, classes with less than 20% completed work (relative to the remaining work) are assigned a red colour. **Perspective 2**. The RC View can also be initiated from the design concept side comprising PBIs (per Feature, Sprint, or Release), enabling users to monitor progress from the perspective of the selected PBI. In this perspective, a user can select an individual Feature from the PBI explorer and invoke the RC View mode. In this case, development progress is calculated and displayed for the particular classes involved in that selected Feature only and using work entry records from that Feature object only (as opposed to work entries from all related PBIs in the first perspective). This second perspective provides another informative RC view from the point view of functional requirement units (a single Feature, a Sprint, or an entire Release) where development progress is then observed for all relevant classes distributed across the city. That is, progress made in relation to the implementation of particular requirements is displayed to the user with a sense of locality across the system.

*4.4.2. Code artefacts-to-glyph mapping*

**Metaphor.** While the mappings discussed in the above section are seen as comprising a novel contribution, the mapping at the level of *code artefacts to visual glyphs* is an established and extensively researched aspect of SV research known as 'metaphor mapping' and which our work simply builds on, as has been described in section 4. A. Specifically, we adopt a very similar metaphor to that of CodeCity and implement the layout algorithm described in [27].

**Glyph Shapes.** In this selected metaphor, the hierarchical structure of software code artefacts is mapped to a corresponding hierarchy of cuboid glyphs such that the whole structure ultimately resembles a city. Packages or modules are mapped to flat rectangular cuboids that work as platforms or districts on which classes are placed and laid out in the form of cuboids resembling buildings. The metaphor we use differs from that of Wettel in that; *a)* we go further by mapping *methods* to cubes contained inside of the class cuboids, and *b)* we distinguish interfaces from classes by mapping them to cylinders instead of cuboids (*this approach has also appeared in* [28]).

**Glyph Dimensions.** In SV it is conventional to map key software metrics to distinctive visual attributes of the adopted metaphor in order to produce a visualisation that provides the most information-rich impression of the actual but intangible software artefacts being visualised (*a principal element of SV theory that aims to scaffold users' spatial ability*). Those typical key metrics are *NOM* (number of methods), *NOA* (number of attributes), and *LOC* (lines of code) for the classes, and *NL* (nesting level) for packages. In Wettel's original metaphor, NOM is mapped to a class glyph's height while NOA is mapped to both width and length of the class glyph. LOC and NL are then mapped to gradient colours of the class and package glyphs, respectively. In ScrumCity the height of a class glyph is mapped to its NOM, but its width and length (which are always equal) are determined by a calculation method consisting of a combination of the class' NOA and NOM which has been found to result in a normalised city landscape that is more realistic (described in more detail in section 7.1). For LOC and NL we adopt the same mapping approach as that of CodeCity, albeit using variant colour gradients (see Fig. [1c]). As for method glyphs, they are always mapped to small cubes with a fixed dimension of one unit and with a default blue colour.

**4.5. Data Collection**

To construct the presented *Conceptual Visualisation*, ScrumCity consumes two categories of raw data, as the above mapping presentation may have readily suggested. These are the *source code* of the system to be visualised and its *Scrum artefact data*. We now discuss each in turn.

**Source Code.** ScrumCity has been developed as an Eclipse plugin to promote accessibility to users, and as a proof of concept it can currently parse Java programs only. Users are thus required to have the ScrumCity plugin installed to their Eclipse platform and to have the system that is to be visualised available as a Java Project in their Eclipse workspace. Being an Eclipse plugin and by virtue of extending the host plugin Vera, ScumCity needs to have ready access to a modelled hierarchical structure of the source code of the target system represented by a single Java object called 'project model repository' (*See section 5.1 for further detail*). Through this repository object, the entire structure of the target system becomes available to ScrumCity in the form of a hierarchy tree, where each code artefact (a method, a class, or a package) can be accessed along with its meta-data, such as LOC, NOA, and NOM, as well as other information. So, in short, once the ScrumCity visualiser is invoked on a particular Java project in the Eclipse workspace, the source code of the target system will be parsed to produce a project's model repository that is then used to create the visualisation glyphs.

**Scrum Artefact Data.** This data mainly represents the standard information pertaining to a typical Scrum development setup, which can be summarised as the list and all meta-data of the Product Backlog Items, Sprints, and Releases of a certain development project, as well as information about their enactment such as estimates, durations, developers' names, priorities, and the like (section 4.3 explains the specific data items required by ScrumCity). As discussed above, this work presumes that developers are using an agile management tool that captures the aforementioned information and provides user dashboards through which developers regularly update their work progress. Other than those typically available data, ScrumCity requires one additional data point to be recorded by developers when updating their dashboard cards (PBI cards), which is designating using QNames the particular code artefacts that were created or modified as a result of the implementation of each PBI. Ideally, all of this information would be captured by the agile management tool (*as most already do except for the QNames trace links*) and can later be exported to XML format. The current implementation of ScrumCity will accept the Scrum Artefact data in XML format files as per the earlier presented schema (multiple files can be fed simultaneously). In future developments, the collection of QName trace links would be semi-automated via an integration mechanism within Eclipse or as an extension to some of the popular agile management tools, such that minimal user effort would be required.

## 5. PROTOTYPING OF SCRUMCITY

### 5.1. Tool building

In order to investigate the feasibility of the proposed Conceptual Visualisation approach, a working prototype tool named ScrumCity, that implemented all the features and functionalities described in this paper, has been developed. Our tool has been inspired by CodeCity, and as discussed above, it implements a very similar approach to that of Wettel's City Metaphor.

While ScrumCity in itself is an independent implementation, rather than re-inventing the wheel entirely it instead builds on earlier research and makes use of existing technologies made available by other researchers. We thus extended the pre-existing Vera Eclipse plugin [30] which was specifically designed to work as a host plugin for other SV tools. The Vera plugin itself utilises the language-independent FAMIX modelling framework to build a project's hierarchical containment model. It first uses Eclipse's native Java parsing capabilities to build an Abstract Syntax Tree (AST) structure, and then feeds it to another importer that creates the final FAMIX model of the project that is being parsed. Fig. [2a] depicts the overall architecture of ScrumCity. For the 3D scene generation and rendering, the API library of jMonkeyEngine3 (jME3) is used along with the third-party Nifty (v.1.3.1) library for creating the embedded GUI controls inside the 3D scene.

### 5.2. Proof of Concept

To test and demonstrate the various capabilities and potential uses of ScrumCity, an initial laboratory assessment was carried out by applying the approach to six real-world open-source systems of varying sizes (the largest having over 1300 classes) that served to verify and validate the tool and its proclaimed features. The subject systems included jME3, ScrumCity, Apache IvyDE, AntViz, jEdit, and Shrimp Suite (the last two being popular SV test subjects). Since obtaining real Scrum data was not a viable option for these systems, a special mechanism was employed in ScrumCity to optionally allow for the generation of simulated Scrum data (for testing purposes) that are specific to the system being visualised when no XML Scrum data file is available.

Fig. [1d] and Fig. [3f] shows jEdit being visualised in ScrumCity. In Fig. [3f], the on-demand GUI controls appear on the right side with the search menu on top and the PBI explorer (Scrum Features) control below it[4]. A distinctive set of red-glowing buildings (system artefacts) stands out prominently amongst the rest in the city of jEdit. This set of buildings (classes and methods) are the specific system components that are implementing the selected simulated sprint. A similar effect results when selecting an individual feature, or even an entire release. Since a given system component can be related to multiple user stories, selecting a building (i.e., a system artefact) will reveal all the features related to it in the list. Two buildings appearing in RC View can also be seen in the centre of Fig. [3f] (buildings with partial orange and sky-blue colours). Figure [3e] shows the entire Shrimp Suite system being visualised and displayed in RC View. Figures [3c] and [3d] show an example of the in situ overlay popup screens displaying textual information, and contextual menus respectively. Fig. [3a] shows the searching functionality with a method being located from within its class.

### 5.3. Presentation Layer

The key objective of the Conceptual Visualisation technique described here is the presentation of aspects of the development process and its activities in a visualisation scene, synchronised alongside the implemented code artefacts, in order to enable stakeholders to review, explore, and reason about the design *in the context of* the actual implementation. Software development processes, however, are typically represented by textual data. Hence, a further challenge encountered in this work relates to how such textual data should be presented *in situ* inside the 3D visualisation scene using a non-intrusive and non-cluttering mechanism. To achieve this, custom-built interactive overlay GUI controls were designed to display the textual information on demand *inside* the 3D scene with minimal intended distraction to the user. These GUI controls include a list interface for displaying the Scrum artefacts (*also referred to as the PBI Explorer in this paper*), a system artefact search menu (Fig. [3a]), a contextual right-click menu (Fig. [3d]), and a multipurpose tooltip menu that can be customized in real time using different keyboard keys to display various sets of information, or even toggled on and off (see Fig. [3f]). *To minimise user distraction and reduce clutter (a critical issue in 3D SV tools), each single GUI control can be instantly shown or hidden with a single keyboard press.*

---

[4] Nifty library provides a tree-list control but due to it having bug issues at present, a simple list is used to demonstrate the concept.

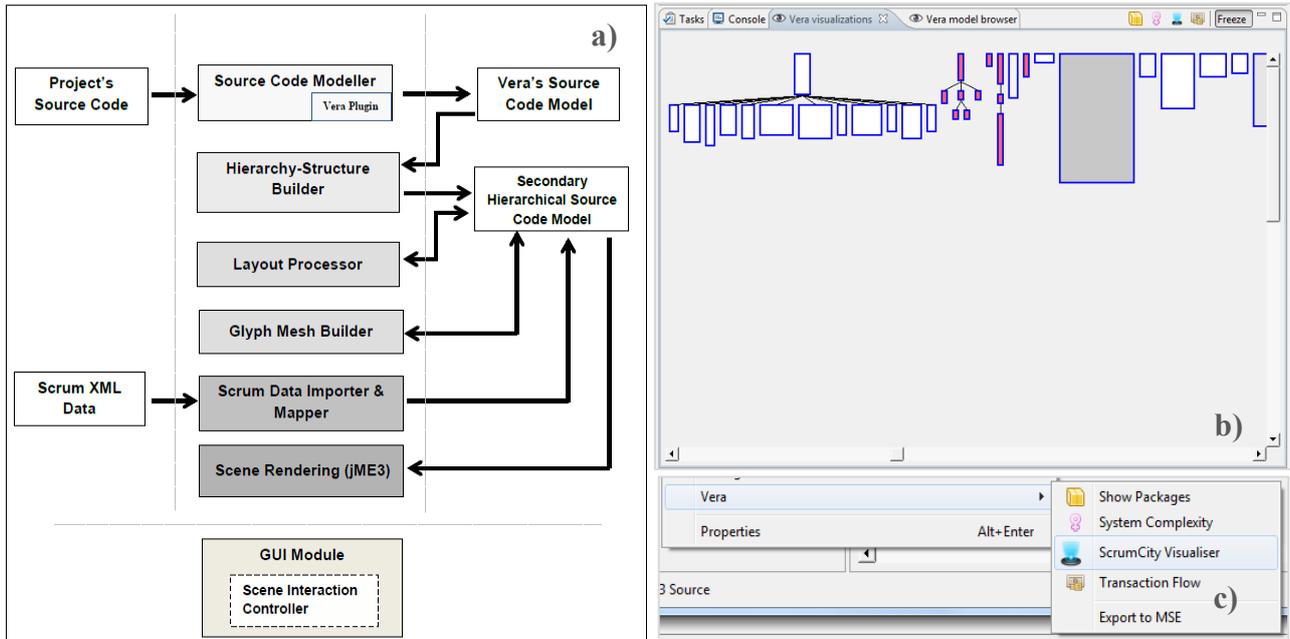

Fig. 2. Clockwise from left, **a)** ScrumCity's Overall Architecture Model, **b)** Vera's Eclipse Plugin showing ScrumCity's Toolbar command with Vera's native visualisation shown in canvas, and **c)** Vera's Contextual Menu (displayed when a *Java* Project is selected in Eclipse's *'Package Explorer'* view).

The key functionalities and features enabled by the presented visualisation approach, and which have been implemented in the prototype tool, are described as follows.

### 5.3.1. Scrum-to-Code Artefact Mapping
We consider this to be the most important functionality delivered by our Conceptual Visualisation. A user can choose to select a particular feature (or PBI) from the *Scrum Features* explorer (*which can be displayed/hidden on demand via a designated keyboard key*) to immediately observe the related code components – the classes and methods – being highlighted in the 3D scene in distinctive red-glowing colour. A class glyph is automatically turned transparent to allow any highlighted method(s) inside to be made visible (see Fig. [3f]). A sprint or release can also be selected to similarly examine the exact system artefacts involved in that sprint or release, revealing the locality of contribution of the selected release or sprint. The Scrum-to-code artefact mapping is bidirectional, so a user can alternatively select a 3D glyph (i.e., a building) of a particular class or method of interest in order to see the related features highlighted in the Scrum Features GUI control, which provides a different perspective. The most prominent advantage of this bidirectional mapping mechanism is that it enables largely effortless *requirements traceability* and *feature location* in an intuitive manner.

### 5.3.2. On-Demand Method View and Interaction
By default, methods are not displayed in the visualised Scrum city to reduce cluttering. Instead, methods become instantly visible when the navigating user comes close to a class by a distance calculated with respect to the size of the class' glyph, achieving '*on-demand detail*'. If the user moves even closer, the class glyph gets removed completely allowing the user to interact directly with the individual methods to view their information and to benefit from traceability at finer granularity. The class glyph returns automatically to its normal appearance upon the user moving away. This feature is achieved via on-demand transparency (*inspired by the earlier work of Balzer et al.* [25], [31]) and on-demand glyph detachment that were both custom-built for the class glyphs. The user also has the option to select any number of class glyphs and invoke a command to instantly reveal their inner methods irrespective of their proximity. Fig. [3b] demonstrates these functionalities.

### 5.3.3. Contextual In Situ Information
A key driver behind our development of Conceptual Visualisation, as discussed above, is that it should enable the provision of information that is considered important to developers (i.e., design concepts and development activities) *where* and *when* they are needed. To facilitate this, once a particular item has been selected in the PBI explorer, the user can then access and read on the spot all the textual information related to the selected Scrum artefact.

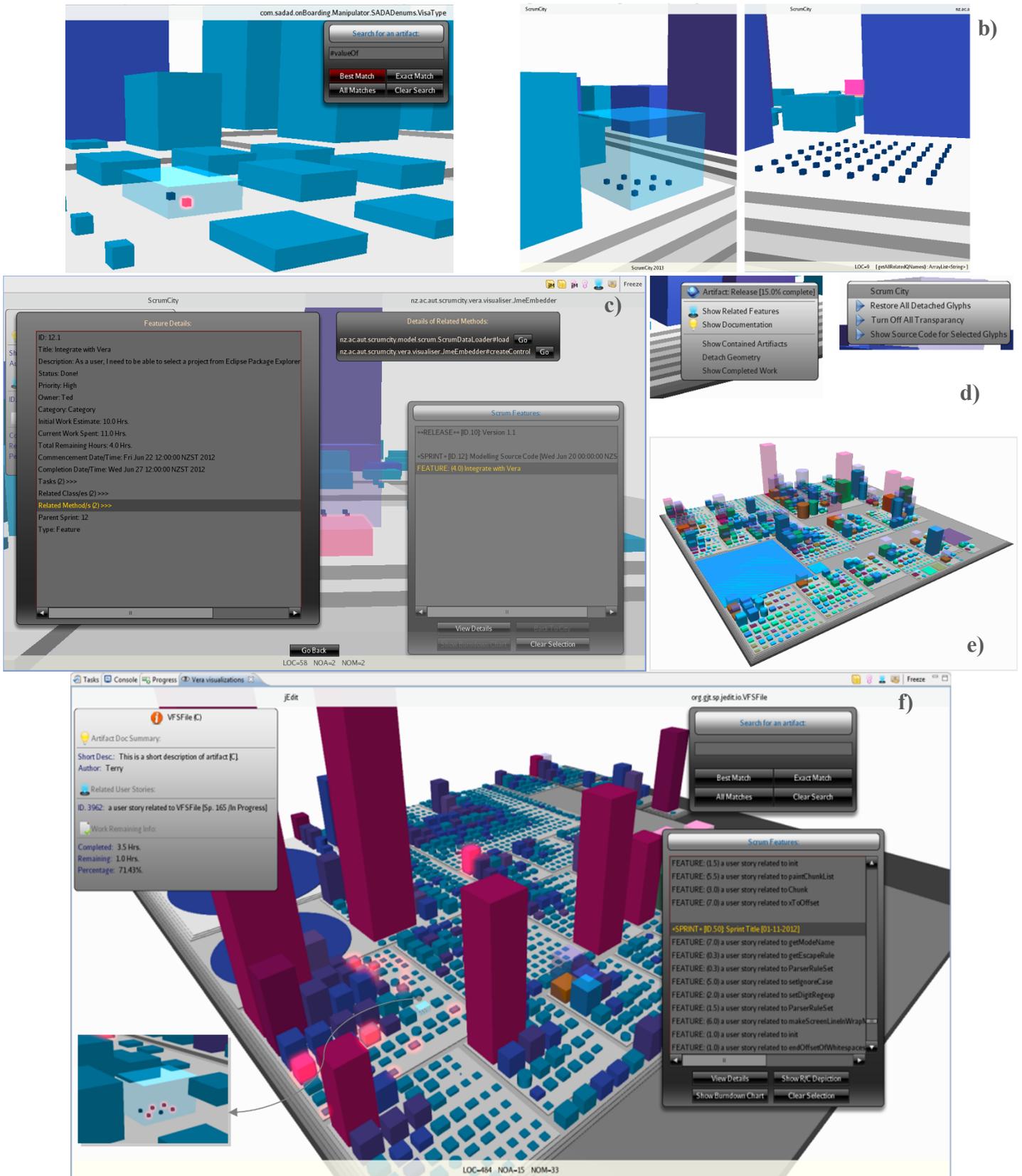

Fig. 3. From left to right: **a)** searching functionality with a method being located from within its class, **b)** on-demand detail: a class glyph is turned transparent as user approaches with the glyph completely detached upon yet further approach to allow interaction with inner methods, **c)** custom in-situ user interfaces in action displaying textual information, **d)** contextual right-click menus, **e)** Shrimp Suite system in simulated RC View (in this view RC colour mappings are applicable), **f)** ScrumCity visualising jEdit system with key features being demonstrated (search tool, PBI explorer, individual classes in RC View in the middle, multipurpose tooltip menu, method views, bidirectional traceability, top and bottom info bars showing artefact names and metrics).

Interactive and non-intrusive overlay popup screens are used to provide easy access to this information in context, without requiring the user to switch to different windows, hence helping to maintain the user's mental focus and avoiding unneeded distraction (see Fig. [3c]). Alternatively, by right-clicking on a system artefact in the 3D scene, a popup menu is displayed providing different contextual commands, including access to the overlay popup screens that display the features related to that specific artefact and all their data. Further, the multipurpose tooltip provides ready access to summarised information about the artefact to which the mouse is pointing (Fig. [3d and 3f]).

### 5.3.4. System Artefact Search
In real-world, large-scale software visualisation, finding a specific artefact of interest manually is challenging; hence, to render our visualisation more effective, a search menu can be summoned on demand (by pressing a designated keyboard key), enabling the user to look up a system artefact by its name. Once found, the artefact is highlighted and the user is automatically transported to it using an animated transition mechanism. According to the literature [32], use of such a transition has significant cognitive advantage over abruptly changing the 3D scene to show the search-result artefact, as it helps the user to retain a mental picture of the new artefact's location relative to other artefacts. Different searching strategies have been implemented giving the user the ability to search for an exact match (leading to a single result) or all available matches (leading potentially to multiple results).

### 5.3.5. Remaining and Completed Work View
By having the system artefacts and their related development process data integrated and synchronised in one scene, a range of potentially useful knowledge can be processed and made readily available to the user, hence informing different software tasks. One such capability is the Remaining/Completed Work view (or *RC View*) which has been largely explained in second paragraph of section 4.4.1. This view is conceived to be potentially useful to managers in particular (or, given the Scrum-centric nature of the approach, for Scrum Masters), for convenient visual monitoring of development progress. Figures [3e] and [3f] demonstrate two modes of this view; two individual classes appear in RC View in the middle of [3f].

## 6. PRELIMINARY EVALUATION
### 6.1. Usage Scenarios

The principal drive behind our proposed Conceptual Visualisation is to support stakeholders in exploring, inspecting, and reasoning about software systems by placing design concepts and their development activities in context with the produced system artefacts. This section describes some illustrative examples of usage scenarios that are enabled by the proposed visualisation. All of the scenarios assume that stakeholders are using the Eclipse platform with the ScrumCity plugin installed. They further assume that Scrum artefact files are loaded into the project.

**Scenario 1.** This scenario demonstrates how refactoring activities can benefit from use of the tool by enabling developers to quickly identify and inspect specific features that might be impacted by their modification of a method of or a class. Given an existing class C1 that dates back to earlier versions of a system, a developer D1 who is tasked with refactoring the code for quality purposes needs to quickly identify its original intended functionality. D1 selects their project folder within Eclipse and invokes the 'ScrumCity Visualiser' command from the contextual menu. D1 is presented with an interactive 3D visualisation scene depicting the entire project as a city. Using the search field of ScrumCity, D1 enters C1's name and, upon invoking the command, C1's glyph (building) is immediately illuminated in a distinctive glowing red colour and the user is transported automatically to that building. D1 selects the building which results in all related features, or Product Backlog Items (PBIs), to be highlighted in the PBI Explorer on the right of the visualisation scene. D1 can now choose one of the highlighted PBIs and invoke the *Details* pane, allowing her to inspect and study the feature's description and its development activity information (see Fig. [3c]). The *Details* pane can include anchors to other classes or methods that are relevant to the selected PBI (if any) which the developer can review to further assess C1's impact on them.

**Scenario 2.** A developer D2 needs to implement some new functionality that requires him to modify method M1. D2 needs to quickly identify the possible impacts on the existing system's functionality before attempting to modify M1. Using similar steps as in Scenario 1, D2 clicks on the method's glyph presented to him in the scene which instantly reveals in the PBI Explorer the particular PBI (or PBIs) that M1 implements. D2 can now inspect the affected functionalities before attempting to modify M1.

**Scenario 3.** This scenario presents a view that is expected to help a user decide if certain refactoring activities might be favoured, through the visual highlighting of abnormalities in implementation. A quality engineer Q1 would like to identify and assess the contribution of Sprint S1 to the system. Q1 selects the S1 tree node from the PBI Explorer which results in various buildings of the city to be immediately illuminated. Q1 can now see a global perspective of S1's implementation and how it is distributed across her software city, hence better informing her decisions (see Figs. [3f and 4b]). For example, this view can reveal to her that S1's implementation is agglomerated in two particular modules, whereas, according to her knowledge of the functionalities being delivered by this sprint, she would instead have expected the implementation to be structured across four specific modules. Another example can be demonstrated at the individual PBI level. Suppose a specific PBI requires database access and, by selecting it in the PBI explorer, the engineer fails to see any illumination in the specialized DB module. This could indicate that the developer had potentially accessed the database directly instead of utilising the specialised DB module – hence a refactoring would be needed to restore the quality and integrity of the codebase. Such a view can be generated for multiple Sprints at once, for multiple features (or PBIs), or even an entire release.

**Scenario 4.** A manager G1 would like to inspect which parts of the system are being modified by his team while they are working on Sprint S2. Following similar steps as in Scenario 3, G1 selects the S2 node in the PBI Explorer allowing him to identify the exact code artefacts that are being modified as a result of S2 (as these are illuminated in red). Further, he can invoke the *Remaining/Completed Work View* to inspect the progress of S2's implementation.

**Scenario 5.** This scenario demonstrates the well-known *Feature Locality* problem, wherein Conceptual Visualisation makes its fulfilment straightforward. A developer D3 is tasked with enhancing certain functionality F1 of a system. D3 thus first wishes to identify the code artefacts contributing to the implementation of F1. D3 locates 2 PBIs in the PBI Explorer that together cover the functionality she needs to enhance[5]. D3 selects the two PBIs simultaneously which results in particular buildings to light up in the city. D3 now has an illuminated view of where exactly F1 is implemented in the system (*in what module(s) and in which classes*) and can also inspect the textual details and description of each PBI right in the context of the visualised system. She can point to any building to see its full Qname in the upper title bar and can choose to navigate directly to the code all within Eclipse. If individual methods rather than an entire class were attributed to one of those PBIs, then their class glyph would turn to transparent, allowing the individual method glyphs inside to shine through their class building. D3 has now identified all the particular code artefacts that she needs to study before adding the new enhancement F1.

**Scenario 6.** This scenario demonstrates Conceptual Visualisation's potential benefit in supporting team and stakeholder communications. A Scrum Master SM1 is chairing a Sprint Review event in which the development team (or teams in the case of parallel Sprints), stakeholders, and the Product Owner are participating. A key element of such an event is to report on sprint progress and to note the product backlog items that were or were not completed. SM1 launches ScrumCity and presents the team with a virtual city depicting their system in its latest status. She further compares this to a screen capture of the system from the last Sprint Review event, offering the participants a visual reference of the latest progress (*which now shows some larger buildings and some other new buildings*). She then selects the particular sprint tree node (or tree nodes in the case of multiple sprints) in the PBI Explorer, resulting in the instant highlighting of all buildings (code artefacts) that were produced or impacted by this sprint; giving the whole team an overview of the scale and 'locality' of this sprint in the wider city code landscape. SM1 now activates the *RC View* for the particular selected sprint, which transforms the impacted buildings into partially filled buildings according to the ratio of work completed vs work remaining for each one, giving a real-time visual impression of the actual progress of the Sprint under review – a view that is localised and fine-grained per code artefact. SM1 is now equipped with a current visualisation of system artefacts through which she can convey the latest status of the product and put such discussion into context; she can designate where this sprint has contributed, she can compare against an older view of the system, and she can display the sprint progress projected against each code artefact (via class buildings).

### 6.2. Preliminary Case Study

The conduct of a full case study would require the availability of a real-world data set consisting of a program's source code, the related Scrum artefact data, their enactment information, work effort and progress information, as well as the QName trace links connecting the Scrum artefacts to their code artefact realisations. Given these demanding requirements and the current preliminary stage of this work, the most promising data set that could be identified is the recent Cassandra artefact produced by Rahimi and Cleland-Huang [33]. Their data set contained 48 feature requirements traced over 27 versions of Cassandra, with class level traceability links identified for each of the 48 features.

Version 2.2.0 in particular (the latest in the data set) was used as a preliminary case study to investigate the feasibility of ScrumCity. It contained a total of 68 packages with 3214 classes. While the feature requirements of the data set are not representing Scrum artefacts *per se*, they still captured the two essential and key elements required by ScrumCity, namely the *small units of functional requirements* (features) augmented by *links* tracing each of those features to their eventual code artefacts. The complete 48 features and their trace links were converted to XML format in accordance with ScrumCity's defined schema, and were represented as individual features (PBIs) of a single Sprint. While we acknowledge this does not reflect the original development methodology used, it was still deemed to be a sensible approximation given that those 48 features represented the actual requirements of Cassandra's initial version 1.0.0-beta1.

Figure [4a] shows the resulting visualisation of Cassandra's v2.2.0, with the entire 48 features of a single sprint selected in the PBI explorer. The first prominent characteristic that is immediately evident is the locality of those original requirements made visible across the system. The user gets immediate feedback on how these old features (*corresponding to version 1.0.0-beta1*) are now distributed across the latest version of the system. Such a view could potentially inform design and quality-related decisions and could be particularly valuable for Sprint Review sessions (see the next section).

Other usage scenarios inspired from the previous section were tested resulting in some interesting findings, and these have showed early promise in terms of delivering real benefits to practitioners. For instance, using the visualisation we were able to immediately identify the exact code artefacts that were implementing the basic 'Caching' functionality (labelled F2). By inspecting the PBI explorer, the user was able to readily locate the Caching feature, and by selecting it, seven classes were instantly revealed in the city (shown as glowing in red in Fig. [4b]).

---

[5] Due to limitations of the Nifty GUI module that was used in the prototype, searching functionality in the PBI Explorer could not be implemented. However, this is a technical issue that should not affect understanding the proposed concept. In a fully developed tool, textual search in the PBI Explorer would be offered.

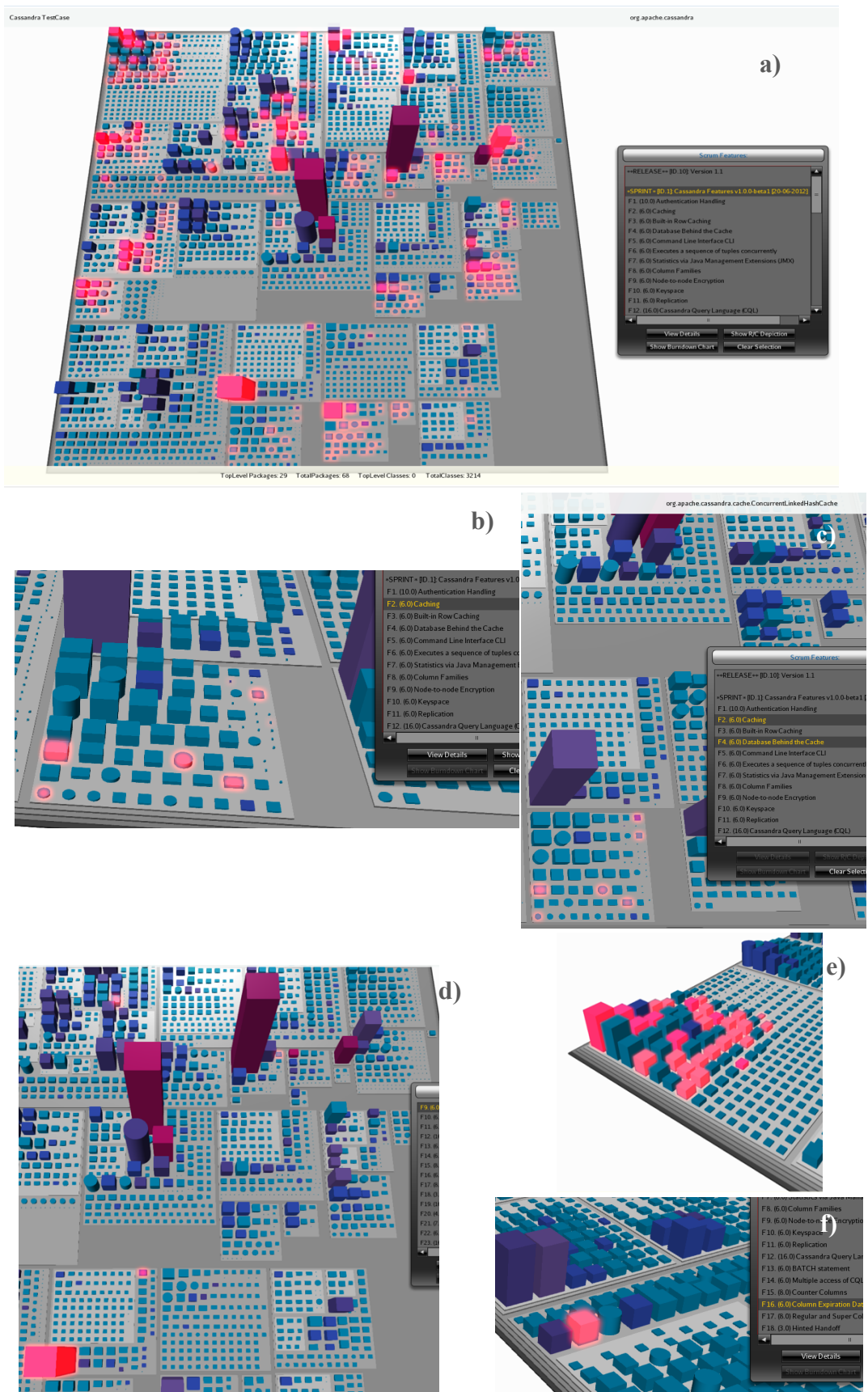

Fig. 4. Left to right from top: **a)** Cassandra's original 48 features (grouped in a sprint in the PBI explorer) demonstrating their implementation locality as seen across version 2.2.0, **b)** selecting the 'Caching' feature in the explorer instantly reveals the code artefacts implementing it, **c)** selecting the 'ConcurrentLinkedHashCache' class in the city reveals two related features in the explorer, **d)** dependency and coupling can be identified in the implementation of feature 9 (Node-to-node Encryption), **e)** complexity (and consequently maintenance effort) is easily identifiable for feature 25 implemented by 29 classes, and for feature 16 that is implemented by a single class (**f**).

If a developer needed to debug or add an enhancement to this functionality, she would now know exactly what class needs to be inspected before carrying out her changes. A team leader or an analyst can also immediately note that the implementation is isolated in a single small package with small number and size of classes, leading to the conclusion that it is of low complexity and should hence require relatively low effort to enhance or debug. Further, if the developer decides that she needs to modify the *ConcurrentLinkedHashCache* class, she can first check if this change could impact other functionalities by simply selecting this class in the city, which would reveal two features being selected in the PBI explorer[6] (see fig [4c]). In addition to the basic caching functionality, the developer can now realise that this particular class is also part of F4 (Database Behind the cache) implementation and thus care must be taken to ensure modifications do not impact this second feature. In comparison, other features such as F16 (Expiring Column) shown in figure [4f] as well as F42 (Ec2 Multi Region Snitch) (not shown) are quickly found to be very simple functionalities that are each realised by a single and independent small class that do not impact any other class or feature.

In a similar mechanism, an analyst investigating the 'Marshalling' functionality (F25) can quickly and easily realise the complexity of this feature (*and hence the effort that would be required to maintain it*) as shown in fig. [4e]. An interesting finding is also observed in feature 9 (Node-to-Node Encryption) where implementation is realised across three modules, possibly an indication of a potential interdependency or coupling (see fig. [4d]).

While the above findings serve to demonstrate the key functionalities of the proposed visualisation, other features of ScrumCity such as the RC View and method level traceability could not be tested as the Cassandra data set did not contain the necessary data. For the future fully developed version, a comprehensive data set would be sought as part of a long-term collaboration with a Scrum practicing team such that a complete set of required data is collected throughout the period.

### 6.3. Preliminary Usability Evaluation

To further investigate the feasibility and the benefits of our approach in the real world, and in order to better inform our future work towards a fully developed tool, we designed a range of usage scenarios based on common development problems, using the Cassandra data set. We used ScrumCity to present a demonstration of the tool to a total of six practising developers with industrial experience ranging from 1 to 8 years. The intent was to have an early and preliminary evaluation of the concept and to receive initial feedback from the industry to guide further development of the work. The problem scenarios that were carried out were similar in nature to the examples discussed previously in Section 6.1. Feedback was collected via a questionnaire of 13 Likert scale questions, and a general comment box. A positive finding of this preliminary evaluation was the unanimous agreement that the tool and the concept behind it address several different practical problems that the developers could relate to. Not just that the developers expressed their approval of its benefits to them, but some cited real scenarios they had recently encountered and related how they could see how the tool could have helped them with these. For instance, two developers from two different organisations indicated that it took extensive efforts for them to locate where particular functionality was implemented in some legacy system they were required to maintain. Given the availability of such a tool, they proclaimed that their efforts could have been substantially reduced. Another developer commented that the view (fig. [4a]) showing the distribution and locality of the entire Sprint implementation (obtained by selecting a Sprint node in the PBI explorer) would address the problem he has recently been tasked with by his manager in their recent Sprint Review session. Other developers highlighted that the tool is potentially valuable to product owners, business analysts, quality engineers, as well as management, and that it would support a range of activities including regression testing, planning, estimation, and risk assessment (for change impacts).

In addition, during the evaluations, several concerns were raised, as well as advice for particular issues to be addressed. The first concern highlighted was the importance of keeping the required effort of QName tagging by developers to a minimum. While some acknowledged that the benefits brought about by the tool would strongly outweigh the effort of a developer having to tag their feature card with a few class or method names, they still contended that full or even semi-automation of this process will have high impact on the tool's adoption by industry. Some expressed that collection of this data point at the IDE side would be highly desirable so that developers perform the tagging as part of their repository commit actions, possibly supported by auto-suggestions. Other concerns that were expressed mostly revolved around usability issues and additional desired features. This included the need to have better focused views in both the city visualisation as well as the PBI explorer. For example, once a feature is selected then all non-relevant code artefacts and features should be hidden or greyed out, as well as the ability to select and work on a particular set of modules rather than having the entire system displayed at all times. Other comments referred to performance improvement, the ability to switch to 2D view, an ability to group classes or modules by dependencies, grouping features by epics, and having the tool as a standalone application, and finally some were concerned about the maximum city size the tool could generate and the practicality when the city becomes extensively large.

Overall this evaluation provided confidence that the concept and tool implementation design have utility for practitioners and is worth refining through further development. The evaluators' ideas for improvement also provided valuable suggestions for future development of the tool.

---

[6] selecting a class would automatically highlight all related classes and all related features

# 7. REFLECTIONS

## 7.1. Metaphor Enhancement

An immediately prominent feature of the city metaphor as employed in ScrumCity that differentiates it from the original Wettel and Lanza [22] metaphor is the representation of methods inside their classes. The representation of methods is particularly important to fully achieve the objective of the presented

Conceptual Visualisation. However, the act of presenting the methods inside their (class) buildings has actually led to some unexpected beneficial results. In spite of the extensive and appropriately positive attention that the Wettel and Lanza metaphor has received from the SV research community, some researchers have highlighted a particular problem with the metaphor in that it can produce software cities with unrealistic landscape appearances [32]; mainly, classes with a high number of methods (NOM) but low number of attributes (NOA) become extremely thin (*needle-like*) while those with low NOM and low NOA become '*dot-like*' making them hardly visible. Apart from some evident disadvantages such as hindering user interactivity in the 3D scene, this great diversity in buildings' shapes also works against the gestalt principle [32]. Most importantly, however, a needle-like building might misrepresent the actual *largeness* of a class and so miscommunicate its importance to those inspecting the code. The introduced modifications to the city metaphor, albeit relatively minor, have resulted unexpectedly in avoiding these irregular shapes and to consistently produce a more uniform city landscape.

## 7.2. Mapping of Scrum Features

It is acknowledged that the collection and availability of the Scrum artefact data (including the QName tagging for each feature or user story) is a key factor for the success and eventual practicality of the presented approach. We contend, however, as discussed earlier, that the majority of the surveyed agile managements tools already capture and make available all the data elements required by ScrumCity, *except for the QName tagging*. This was further confirmed with the developers from industry during the preliminary evaluation. They affirmed also that they are required to regularly update their feature (or user story) cards in their tools' dashboards with progress details (work hours or effort) which would enable the RC View of ScrumCity, albeit they consented that some developers do not necessarily adhere to it consistently. As for the QName tagging, except for one developer, they all expressed their willingness to record this additional data point given its low overhead compared with the considerable advantages it brings to them. Encouraged by the range of solutions and benefits the approach enables, one developer expressed confidence that his management would certainly approve such an additional small requirement to be collected. Nonetheless, it is admitted that seamless integration with existing tools is still a critical factor for the approach's successful adoption by industry, and future work on ScrumCity will focus on integration issues and the automation of QName tagging.

## 7.3. Implications for Practice

The previous section (6) has served to demonstrate a good range of potential practical usages and real-world problems that could be better supported by the tool, with initial feedback from industry that is largely positive. Such uses included scenarios not anticipated earlier by the authors, such as those of regression testing, effort estimation, and risk assessment. While requirement traceability and feature locality were the key supported activities, other uses were in fact indirectly enabled as a consequence of synchronising the original design concepts with their end product code artefacts. This is strongly aligned with the findings of other researchers such as Kuhn et al. [10] in their 2010 work where they emphasised the benefits and importance of making design concepts contextually available to users in a synchronised manner.

# 8. CONCLUSIONS AND FUTURE WORK

This paper has introduced a novel visualisation approach that makes available the conceptual design behind a system and its development activities, which according to the discussed literature should account for a major missing aspect of software in existing SV approaches and that has the potential to make SV tools better equipped to answer important questions that software stakeholders ask in the real world. The feasibility of the approach has been demonstrated through a prototype tool named ScrumCity. This tool has been used to conduct a preliminary case study using a real-world data set that served to demonstrate various usage scenarios in solving real-world problems. As a preliminary evaluation, feedback has also been collected from developers with industrial experience, which has largely showed consistent approval of the potential benefits and advantages for real world applications (*especially that of requirement traceability and feature location*) enabled by the approach, with a relatively wide range of potential users that includes developers, testers, quality engineers, and product owners. The conducted preliminary evaluation has also resulted in valuable feedback that will be vital in informing future development and enhancement of the tool. Through a future collaboration with a Scrum-practicing organization from industry, a full empirical evaluation of the visualisation approach will be carried out to further establish the viability and effectiveness of the concept, that should include other ScrumCity features (such as the RC View) that were not tested at this stage.

## Acknowledgments

The authors gratefully acknowledge the financial support provided to them by the Government of the Kingdom of Saudi Arabia and Auckland University of Technology. They would like to express their gratitude to Mona Rahimi and her collaborators for making available the Cassandra dataset, and for her kind support in answering their enquiries. They would also like to thank the developers who kindly participated in the preliminary evaluation of ScrumCity, and for the valuable feedback and comments that were received. Finally, they would like to acknowledge the effort of the editors and reviewers of the initial versions of this paper, and thank them for the insights and the important feedback they provided, and which has led to significant improvement of this paper.